\documentclass[11pt,english]{scrartcl}
\usepackage{lmodern}
\usepackage[T1]{fontenc}
\usepackage[latin9]{inputenc}
\usepackage{geometry}
\usepackage{amsmath}
\usepackage{amssymb}
\usepackage{amsfonts}
\usepackage{graphicx}
\usepackage{color}
\usepackage{soul} 

\allowdisplaybreaks
\begin{document}
\section*{Comment on ``Identifying Functional Thermodynamics in Autonomous
Maxwellian Ratchets'' (arXiv:1507.01537v2)}

\vspace{1cm}

\begin{center}
Neri Merhav\\
Department of Electrical Engineering\\
Technion -- Israel Institute of Technology\\
Haifa 3200004, Israel\\
{\tt merhav@ee.technion.ac.il}\\
\end{center}

\vspace{1cm}

\begin{center}
We make a few comments on some misleading statements in the above paper.
\end{center}

\vspace{1cm}

The above article is about a family of Maxwell--like demons for which
there are correlations between ``information--bearing degrees of freedom''
(quoting from the authors' description of their work).
More precisely, it is another paper (in quite a long series of recent papers)
about an extended second law for a family of models of
systems that consist of an information reservoir (in the form of a running
digital tape) and it is shown that the amount of heat that can be transformed
into work, done by the system, is upper bounded by the entropy increase
of the information reservoir (in units of $kT$). The difference between 
this work and previous works is claimed be the successful incorporation of possible
statistical dependencies (i.e., correlations) 
between information--bearing degrees of freedom.
One of the claimed main results is that in their more general framework,
the amount of heat transformed into work is upper bounded in terms of the
increase in the {\it joint} entropy rate of the sequence bits after the
interaction, relative to their joint entropy before the interaction
with the demon (see eq.\ (4) in the paper).

I have several comments regarding specific points in this paper.
\begin{enumerate}

\item The first two sentences of the Abstract read as follows:
{\it ``We introduce a family of Maxwellian Demons for which correlations among
information bearing degrees of freedom can be calculated exactly and in
compact analytical form. This allows one to precisely determine Demon
functional thermodynamic operating regimes, when previous methods either
misclassify or simply fail due to the approximations they invoke.''} These
two opening sentences give the reader a very strong misleading impression that
the paper above is the {\bf first} to incorporate correlations successfully in
general. However, in a recent paper [1], which was published before the arXiv post under
discussion (and actually even cited therein -- ref.\ [57]), 
this has already been done for a model in the same spirit
(a simplified version of the Mandal--Jarzynski model). Specifically, in [1]
the extended second law of information thermodynamics was further generalized
in several directions, one of which allows correlations in the incoming
bits of the information sequence (see Section 4 of [1]). The main result in
Section 4 of [1] was {\bf exactly} the same as 
the above mentioned upper bound on the extracted work in terms
of the change in the joint entropy (more details to follow).
Note that in [1], the joint distribution of the input bits was assumed
completely general. Even stationarity was not assumed.

\item On page 2 (left column, second to the last paragraph) of paper 1507.01537v2, there is another
misleading statement: {\it ``In effect, they account for Demon
information--processing by replacing the Shannon information of the components as a whole 
by the sum of the components' individual Shannon informations. Since the
latter is larger than the former [19], these analyses lead to weak bounds on
the Demon performance.''} The last sentence is simply not true in general.
Using the authors' notation, this is true if the incoming bits $\{Y_i\}$ are
statistically independent, while the stochastic transformation (the channel)
from $\{Y_i\}$ to the outgoing bits, $\{Y_i'\}$, is a channel with memory
(i.e., it is not a memoryless channel where $P(Y_1',\ldots,Y_n'|Y_1,\ldots,Y_n)$
factors into a product form $\prod_{i=1}^n P(Y_i'|Y_i)$). However, it may
not be true if the input $\{Y_i\}$ is self--correlated. 
Indeed, in [1], where the input is self--correlated and the channel is
memoryless, it is the other way around: the sum of individual entropy
differences is smaller (and hence tighter) than the joint entropy difference.
The assumption that
the input $\{Y_i\}$ is i.i.d.\ is made much later in 1507.01537v2 (note
that even eq. (3) therein, which comes much later, still gives the impression
that the input is not necessarily i.i.d.). This is why the above cited
sentence is misleading. But beyond all this, the point in the second law and its extensions
should not necessarily be just to bound the amount of the extractable
work.\footnote{In general, bounds are useful when they are easier to calculate than the
real quantity of interest, which is not quite the case in this context. Quite 
the contrary, joint entropies (especially of long blocks) are much harder to
calculate than the work itself, which depends only on the input and output
marginals.}
The point should be to provide, first and foremost, an extended version of the second 
law in a faithful manner, namely, to show the increase of the {\bf real}
entropy of the entire system, including that
of the information reservoir. In the correlated case, the latter is given be
the change in the joint entropy of the symbols, regardless of whether or not
this is smaller or larger than the sum of individual entropy differences.

\item After eq.\ (4), which is claimed to be one of the main results in 1507.01537v2, it says: {\it 
``Other bounds that account for correlations have been analyzed in the context
of a memoryless channel driven by a memoryful process [57].''}
This is yet another misleading sentence. The bound in [57] (which is ref.\ [1] below)
is {\bf exactly} the same as in eq. (4) of 1507.01537v2, except that in [57],
no limit on is taken over the normalized entropies (but this is because
even stationarity is not assumed there, so the limit might not exist).
Moreover, while it is true 
that in the model of [57] the channel was memoryless, the {\it derivation
itself} of this very same bound (in Section 4 of [57]) was not sensitive to the channel
memorylessness assumption. The crucial step in [57] (in the current notation) was
the equality $H(Y_i'|Y_1,\ldots,Y_{i-1},Y_1',\ldots,Y_{i-1}')=
H(Y_i'|Y_1,\ldots,Y_{i-1})$, which is the case when $Y_i'\to
(Y_1,\ldots,Y_{i-1})\to (Y_1',\ldots,Y_{i-1}')$ forms a Markov chain, and this
happens not only for a memoryless channel, but for any causal channel 
without feedback, namely, $P(Y_1',\ldots,Y_n'|Y_1,\ldots,Y_n)=\prod_{i=1}^n
P(Y_i'|Y_1,\ldots,Y_i)$. In physical terms, this actually means full
generality.

\end{enumerate}

\section*{References}

[1] N.~Merhav, ``Sequence complexity and work extraction,''
{\it Journal of Statistical Mechanics: Theory and Experiment},
P06037, June 2015. doi:10.1088/1742-5468/2015/06/P06037

\end{document}